\pdfoutput=1

\documentclass[12pt,a4paper]{article}

\usepackage{ifthen} 
\newboolean{pdflatex}
\setboolean{pdflatex}{true} 

\newboolean{articletitles}
\setboolean{articletitles}{true} 

\newboolean{uprightparticles}
\setboolean{uprightparticles}{false} 

\def\paperauthors{LHCb collaboration} 
\def\paperasciititle{Search for the doubly charmed baryon Omegacc+} 
\def\papertitle{Search for\\ the doubly charmed baryon \Omegacc} 
\def\paperkeywords{{High Energy Physics}, {LHCb}} 
\def\papercopyright{\the\year\ CERN for the benefit of the LHCb collaboration} 
\def\paperlicence{CC BY 4.0 licence}
\def\paperlicenceurl{https://creativecommons.org/licenses/by/4.0/}

\usepackage[top=1in, bottom=1.25in, left=1in, right=1in]{geometry}

\usepackage{microtype}
\usepackage{lineno}  
\usepackage{xspace} 
\usepackage{caption} 

\usepackage{graphicx}  
\usepackage{color}
\usepackage{colortbl}
\graphicspath{{./figs/}} 
\DeclareGraphicsExtensions{.pdf,.PDF,png,.PNG}

\usepackage{amsmath} 
\usepackage{amssymb}
\usepackage{amsfonts}
\usepackage{upgreek} 

\newcommand*\patchAmsMathEnvironmentForLineno[1]{%
\expandafter\let\csname old#1\expandafter\endcsname\csname #1\endcsname
\expandafter\let\csname oldend#1\expandafter\endcsname\csname
end#1\endcsname
 \renewenvironment{#1}%
   {\linenomath\csname old#1\endcsname}%
   {\csname oldend#1\endcsname\endlinenomath}%
}
\newcommand*\patchBothAmsMathEnvironmentsForLineno[1]{%
  \patchAmsMathEnvironmentForLineno{#1}%
  \patchAmsMathEnvironmentForLineno{#1*}%
}
\AtBeginDocument{%
\patchBothAmsMathEnvironmentsForLineno{equation}%
\patchBothAmsMathEnvironmentsForLineno{align}%
\patchBothAmsMathEnvironmentsForLineno{flalign}%
\patchBothAmsMathEnvironmentsForLineno{alignat}%
\patchBothAmsMathEnvironmentsForLineno{gather}%
\patchBothAmsMathEnvironmentsForLineno{multline}%
\patchBothAmsMathEnvironmentsForLineno{eqnarray}%
}

\usepackage{hyperxmp}

\usepackage[pdftex,
            pdfauthor={\paperauthors},
            pdftitle={\paperasciititle},
            pdfkeywords={\paperkeywords},
            pdfcopyright={Copyright (C) \papercopyright},
            pdflicenseurl={\paperlicenceurl}]{hyperref}

\usepackage[colorinlistoftodos,textsize=scriptsize]{todonotes}

\usepackage[all]{hypcap} 

\usepackage{xspace} 
\usepackage{upgreek}

\def\lhcb   {\mbox{LHCb}\xspace}                 
 \def\PXi         {\ensuremath{\Xi}\xspace}                 
 \def\PLambda     {\ensuremath{\Lambda}\xspace}                               
 \def\POmega      {\ensuremath{\Omega}\xspace}                              
 \def\Ppi         {\ensuremath{\pi}\xspace}                                           
 \def\Ppsi        {\ensuremath{\psi}\xspace}                               
 \mathchardef\PDelta="7101
 \mathchardef\PXi="7104
 \mathchardef\PLambda="7103
 \mathchardef\PSigma="7106
 \mathchardef\POmega="710A
 \mathchardef\PUpsilon="7107             
 \def\PB      {\ensuremath{B}\xspace}                                             
 \def\PJ      {\ensuremath{J}\xspace}                 
 \def\PK      {\ensuremath{K}\xspace}                                                        
 \def\Pb      {\ensuremath{b}\xspace}                 
 \def\Pc      {\ensuremath{c}\xspace}                 
 \def\Pd      {\ensuremath{d}\xspace}                 
 \def\Pp      {\ensuremath{p}\xspace}                              
 \def\Ps      {\ensuremath{s}\xspace}                                
 \def\Pu      {\ensuremath{u}\xspace}                 
\def\uquark    {{\ensuremath{\Pu}}\xspace}
\def\dquark    {{\ensuremath{\Pd}}\xspace}
\def\squark    {{\ensuremath{\Ps}}\xspace}
\def\cquark    {{\ensuremath{\Pc}}\xspace}
\def\bquark    {{\ensuremath{\Pb}}\xspace}
\def\pion   {{\ensuremath{\Ppi}}\xspace}
\def\pip    {{\ensuremath{\pion^+}}\xspace}
\def\pim    {{\ensuremath{\pion^-}}\xspace}
\def\kaon    {{\ensuremath{\PK}}\xspace}
\def\Km      {{\ensuremath{\kaon^-}}\xspace}
\def\B       {{\ensuremath{\PB}}\xspace}
\def\Bc      {{\ensuremath{\B_\cquark^+}}\xspace}
\def\jpsi     {{\ensuremath{{\PJ\mskip -3mu/\mskip -2mu\Ppsi\mskip 2mu}}}\xspace}
\def\proton      {{\ensuremath{\Pp}}\xspace}
\def\Lz          {{\ensuremath{\PLambda}}\xspace}
\def\Xires       {{\ensuremath{\PXi}}\xspace}
\def\Omegares    {{\ensuremath{\POmega}}\xspace}
\def\Lc          {{\ensuremath{\Lz^+_\cquark}}\xspace}
\def\Xicp        {{\ensuremath{\Xires^+_\cquark}}\xspace}
\def\Lb           {{\ensuremath{\Lz^0_\bquark}}\xspace}
\def\BF         {{\ensuremath{\mathcal{B}}}\xspace}
 
\def\to                 {\ensuremath{\rightarrow}\xspace}
\def\eps   {{\ensuremath{\varepsilon}}\xspace}
\newcommand{\nospaceunit}[1]{\ensuremath{\text{#1}}}       
\newcommand{\aunit}[1]{\ensuremath{\text{\,#1}}}                   
\newcommand{\tev}{\aunit{Te\kern -0.1em V}\xspace}
\newcommand{\gev}{\aunit{Ge\kern -0.1em V}\xspace}
\newcommand{\mev}{\aunit{Me\kern -0.1em V}\xspace}
\newcommand{\mevc}{\ensuremath{\aunit{Me\kern -0.1em V\!/}c}\xspace}
\newcommand{\gevc}{\ensuremath{\aunit{Ge\kern -0.1em V\!/}c}\xspace}
\newcommand{\mevcc}{\ensuremath{\aunit{Me\kern -0.1em V\!/}c^2}\xspace}
\newcommand{\gevcc}{\ensuremath{\aunit{Ge\kern -0.1em V\!/}c^2}\xspace}
\def\mum  {\ensuremath{\,\upmu\nospaceunit{m}}\xspace}
\def\nb {\aunit{nb}\xspace}

\def\fb   {\ensuremath{\aunit{fb}}\xspace}
\def\invfb   {\ensuremath{\fb^{-1}}\xspace}
\def\fs   {\ensuremath{\aunit{fs}}\xspace}
\newcommand{\chisq}{\ensuremath{\chi^2}\xspace}
\newcommand{\chisqip}{\ensuremath{\chi^2_{\text{IP}}}\xspace}
\def\pt         {\ensuremath{p_{\mathrm{T}}}\xspace}

\def\evtgen     {\mbox{\textsc{EvtGen}}\xspace}
\def\geant      {\mbox{\textsc{Geant4}}\xspace}
\def\photos     {\mbox{\textsc{Photos}}\xspace}
\def\pythia     {\mbox{\textsc{Pythia}}\xspace}
\newcommand{\eg}{\mbox{\itshape e.g.}\xspace}
\def\Xicp         {{\ensuremath{\Xires^+_\cquark}}\xspace}
\def\Xiccpp       {{\ensuremath{\Xires^{++}_{\cquark\cquark}}}\xspace}
\def\Xiccp        {{\ensuremath{\Xires^{+}_{\cquark\cquark}}}\xspace}
\def\genxicctwo   {\mbox{\textsc{GenXicc2.0}}\xspace}

\def\Xicc{{\ensuremath{\Xires_{\cquark\cquark}}}\xspace}
\def\Omegacc{{\ensuremath{\Omegares^{+}_{\cquark\cquark}}}\xspace}

\usepackage{multirow} 
\usepackage{booktabs} 
\usepackage{rotating}

\usepackage{siunitx}
\usepackage{cite} 
\usepackage{mciteplus}

\usepackage{longtable} 

\begin{document}

\renewcommand{\thefootnote}{\fnsymbol{footnote}}
\setcounter{footnote}{1}



\begin{titlepage}
\pagenumbering{roman}

\vspace*{-1.5cm}
\centerline{\large EUROPEAN ORGANIZATION FOR NUCLEAR RESEARCH (CERN)}
\vspace*{1.5cm}
\noindent
\begin{tabular*}{\linewidth}{lc@{\extracolsep{\fill}}r@{\extracolsep{0pt}}}
\ifthenelse{\boolean{pdflatex}}
{\vspace*{-1.5cm}\mbox{\!\!\!\includegraphics[width=.14\textwidth]{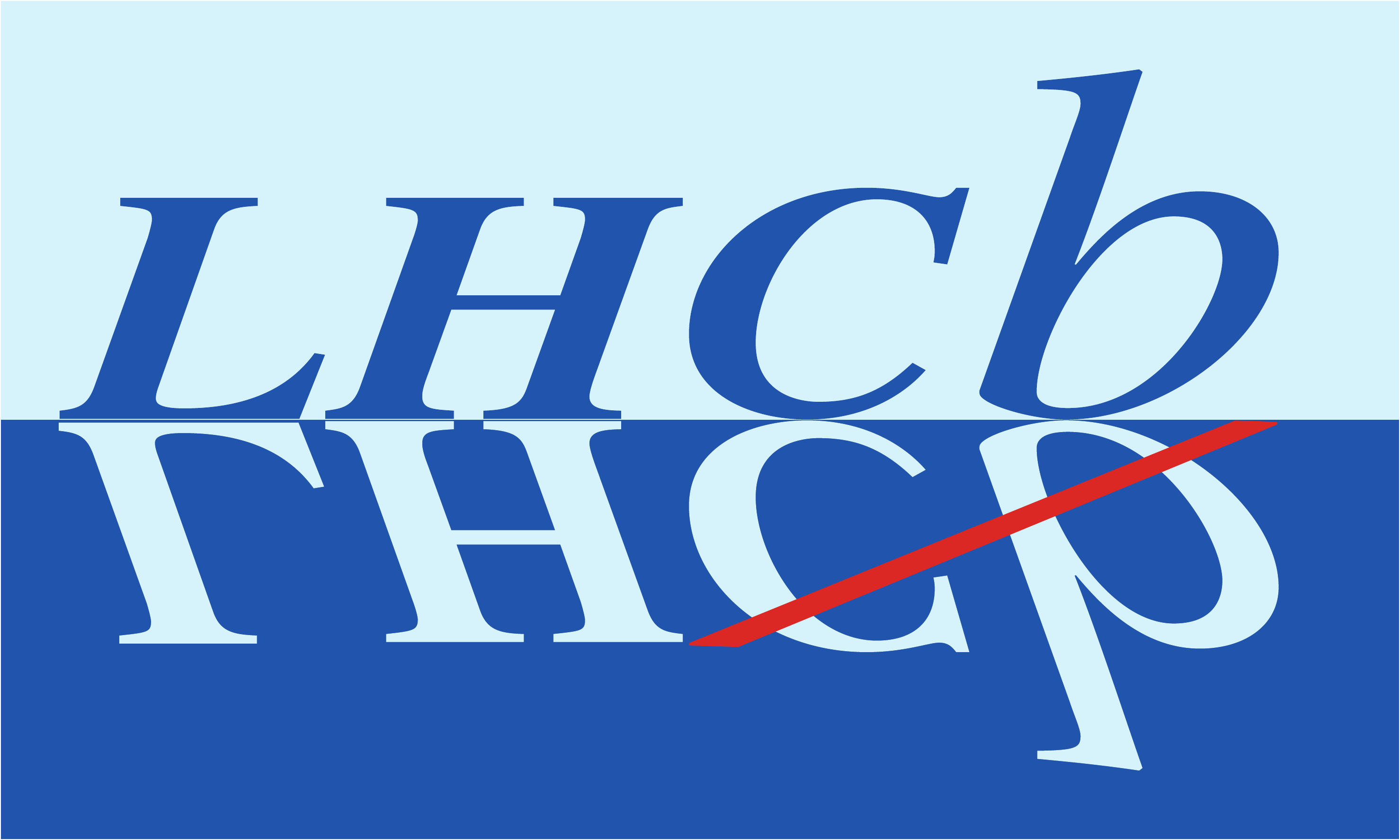}} & &}%
{\vspace*{-1.2cm}\mbox{\!\!\!\includegraphics[width=.12\textwidth]{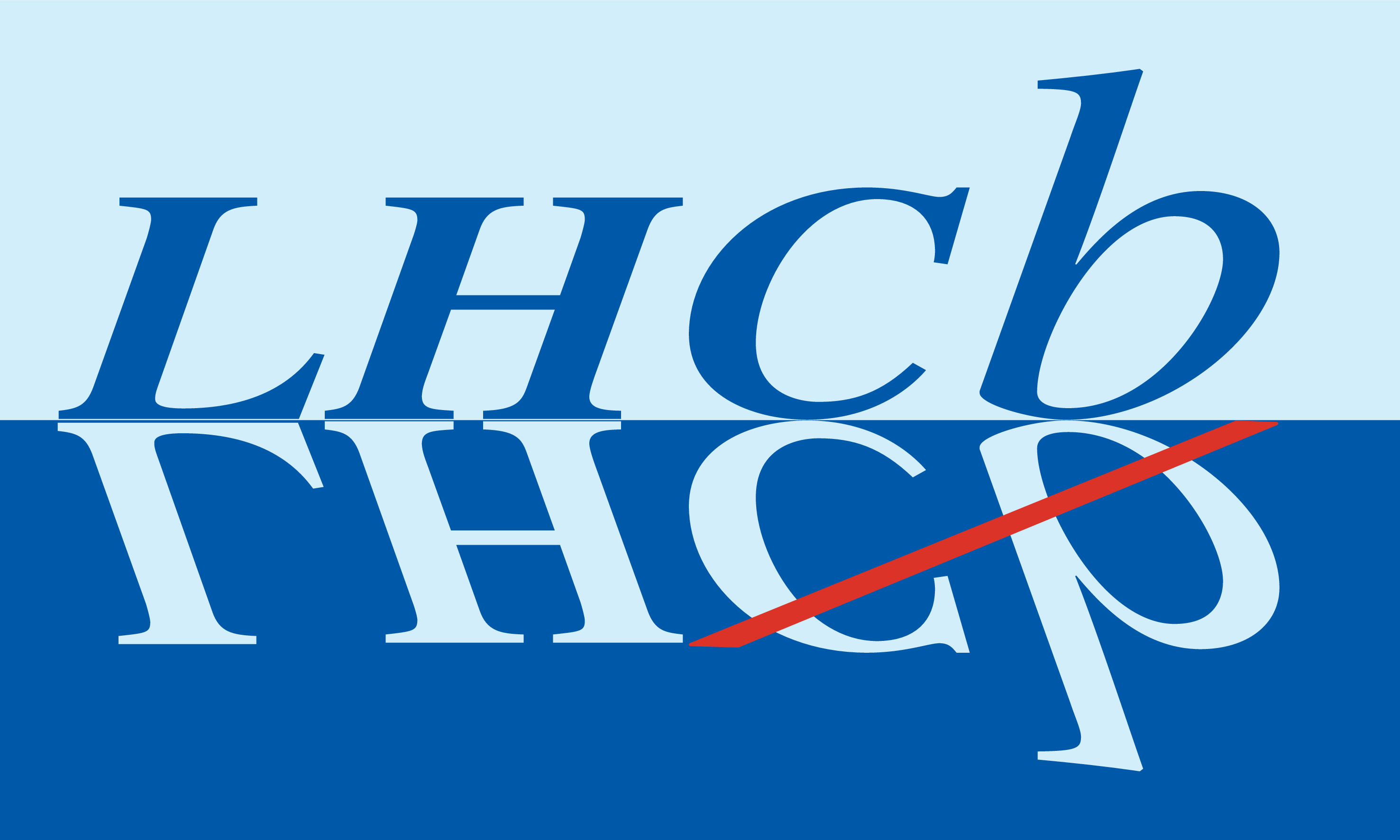}} & &}%
\\
 & & CERN-EP-2021-068 \\  
 & & LHCb-PAPER-2021-011 \\  
 & & \today \\ 
 & & \\
\end{tabular*}

\vspace*{4.0cm}

{\normalfont\bfseries\boldmath\huge
\begin{center}
  \papertitle 
\end{center}
}

\vspace*{2.0cm}

\begin{center}
\paperauthors\footnote{Authors are listed at the end of this paper.}
\end{center}

\vspace{\fill}

\begin{abstract}
 \noindent
  A search for the doubly charmed baryon \Omegacc with the decay mode $\Omegacc\to\Xicp\Km\pip$ is performed 
  using proton-proton collision data at a centre-of-mass energy of
  13\tev collected by the \lhcb experiment from 2016 to 2018,
  corresponding to an integrated luminosity of 5.4\invfb. No significant signal is observed within the invariant mass range of 3.6 to 4.0\gevcc. Upper limits are set on the ratio $R$ of the production cross-section times the total branching fraction of the $\Omegacc\to\Xicp\Km\pip$ decay with respect to the $\Xiccpp\to\Lc\Km\pip\pip$ decay. Upper limits at 95\% credibility level for $R$ in the range  $0.005$ to $0.11$ are obtained for different hypotheses on the \Omegacc mass and lifetime in the rapidity range from 2.0 to 4.5 and transverse momentum range from 4 to 15\gevc.
\end{abstract}

\vspace*{2.0cm}

\begin{center}
  Published in
  Sci. China-Phys. Mech. Astron. 64, 101062 (2021)
\end{center}

\vspace{\fill}

{\footnotesize 
\centerline{\copyright~\papercopyright. \href{\paperlicenceurl}{\paperlicence}.}}
\vspace*{2mm}

\end{titlepage}


\newpage
\setcounter{page}{2}
\mbox{~}
%
%
%
%
\cleardoublepage


\renewcommand{\thefootnote}{\arabic{footnote}}
\setcounter{footnote}{0}



\pagestyle{plain} 
\setcounter{page}{1}
\pagenumbering{arabic}


%

\section{Introduction}%
\label{sec:introduction}

The quark model~\cite{GellMann:1964nj,Zweig:1981pd,Zweig:1964jf} predicts the existence of multiplets of baryon states with a structure containing three valence quarks, two charm quarks and a light quark (\uquark, \dquark or \squark). There are three doubly charmed, weakly decaying states expected: a \Xicc isodoublet $(ccu,ccd)$ and an \Omegacc isosinglet $(ccs)$, each with spin-parity $J^{P} = 1/2^{+}$. Theoretical models~\cite{Fleck:1989mb,Berezhnoy:1998aa, Ebert:2002ig, Chang:2006eu} predict that the light quark moves with a large relative velocity with respect to the bound ($cc$)-diquark inside the baryon and experiences a short-range of QCD potential. 

The \Xiccpp baryon with mass $3620.6 \pm 1.6$\mevcc was first observed by the LHCb collaboration in the \Lc\Km\pip\pip decay\footnote{Inclusion of charge-conjugated processes is implied throughout this paper.}~\cite{LHCb-PAPER-2017-018}, and confirmed in the $\Xicp\pip$ decay~\cite{LHCb-PAPER-2018-026}. The search for \Xiccp via its decay to \Lc\Km\pip was updated recently by the LHCb collaboration, and no significant signal was found~\cite{LHCb-PAPER-2019-029}. The \Omegacc mass is predicted to be in the range  $3.6-3.9\gevcc$~\cite{Ebert:2002ig,Gershtein:2000pd,Ebert:1997pc,Roncaglia1:1995pd,Roncaglia2:1995pd,Korner:1994np,Narodetskii:2001hep,Zachary:2014pd,Kazem:2012,Valcarce:2008,Vijande:2013} and its lifetime is predicted to be $75-180\fs$~\cite{Haiyang:2018pd,hsi2008lifetime, Ebert:2002ig, guberina1999inclusive,Karliner:2014gca,Kiselev:1998sy,Kiselev:2001fw}. 
Due to destructive Pauli interference~\cite{Haiyang:2018pd}, the \Xiccpp and \Omegacc baryons have a larger lifetime than that of the \Xiccp baryon which is shortened by the contribution from 
$W$ boson exchange between the charm and down quarks. In proton-proton ($pp$) collisions at a centre-of-mass energy of 13\tev, the production cross-section of the doubly charmed baryons is predicted to be within the range of $60-1800\nb$~\cite{Berezhnoy:1998aa,Kiselev:2001fw,Ma:2003zk,Chang:2006xp,Chang:2006eu,Zhang:2011hi,Chang:2005bf}, which is between $10^{-4}$ and $10^{-3}$ times that of the total charm quark production~\cite{LHCb-PAPER-2019-035}. 
The production cross-section of the \Omegacc baryon is expected to be about 1/3 of those of the
\Xiccp and \Xiccpp baryons due to the presence of an $\squark$ quark~\cite{Berezhnoy:2018pd}. 
A discovery of the \Omegacc baryon and measurements of its properties would validate the aforementioned theoretical predictions, and deepen our understanding on the dynamics in the production and decays of the doubly charmed baryons.

\begin{figure}[!htp]
  \centering
  \includegraphics[width=0.55\linewidth]{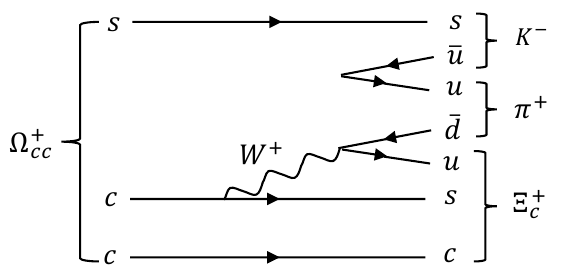}
  \caption{Example Feynman diagram for the $\Omegacc\to\Xicp\Km\pip$ decay.}
  \label{fig:feynman1}
\end{figure}

In this paper, a search for the \Omegacc baryon via the $\Omegacc \to
\Xicp \Km \pip$ decay, which is predicted to have a relatively large branching fraction~\cite{Shi:2017dto,Jiang:2018}, is presented. The data are collected by the LHCb experiment in $pp$ collisions at a centre-of-mass energy of 13\tev in the period from 2016 to 2018. A possible Feynman diagram for this decay
is shown in Fig.~\ref{fig:feynman1}. 

In order to avoid experimenters' bias, the results of the analysis were not examined until the full procedure had been finalised.
Two different selections are developed: selection A is optimised to maximise the hypothetical signal sensitivity and selection B is optimised for the production ratio measurement. 
The analysis strategy is defined as follows:
\mbox{selection A} is first used to search for $\Omegacc$ signal   
and evaluate its significance as a function of $\Omegacc$ mass. 
If evidence for a signal with a global significance above 3 standard deviations after considering the look-elsewhere effect would be found, the mass would be measured and Selection B would be employed to measure the production cross-section of the \Omegacc baryon;
else, upper limits on the production ratio $R$ as a function of the \Omegacc mass for different lifetime hypotheses would be set. The production ratio $R$, relative to the $\Xiccpp \to \Lc\Km\pip\pip$ decay, is defined as
\begin{equation}
  \label{eq:RXicc}
  R \equiv \frac{\sigma(\Omegacc)\times\BF(\Omegacc \to
  \Xicp\Km\pip)\times\BF(\Xicp \to
  \proton\Km\pip)}{\sigma(\Xiccpp)\times\BF(\Xiccpp \to \Lc\Km\pip\pip)\times\BF(\Lc \to
  \proton\Km\pip)},
\end{equation}
where $\sigma$ is the baryon production cross-section and
$\mathcal{B}$ is the branching fraction of the corresponding decays. Both the $\Omegacc$ and $\Xiccpp$ baryons are required to be in the rapidity range of 2.0 to 4.5 and have  transverse momentum between 4 and 15 \gevc.

The production ratio is evaluated as
\begin{equation}
\label{eq:alphaN}
  R = \frac{\varepsilon_{\text{norm}}}{\varepsilon_{\text{sig}}}
                \frac{N_{\text{sig}}}{N_{\text{norm}}}
              \equiv \alpha N_{\text{sig}},
\end{equation}
where $\varepsilon_{\text{sig}}$ and $\varepsilon_{\text{norm}}$ 
refer to the efficiencies of the \Omegacc signal 
and the \Xiccpp normalisation decay mode, respectively,
$N_{\text{sig}}$ and $N_{\text{norm}}$ are the corresponding yields, and $\alpha$
is the single-event sensitivity. 
The lifetime of the \Omegacc baryon is unknown and strongly affects the selection efficiency,
hence upper limits on $R$
are quoted as a function of the \Omegacc baryon mass  
for a discrete set of lifetime hypotheses.

\section{Detector and simulation}%
\label{sec:Detector}

The \lhcb detector~\cite{LHCb-DP-2008-001,LHCb-DP-2014-002} is 
a single-arm forward
spectrometer covering the \mbox{pseudorapidity} range $2<\eta <5$,
designed for the study of particles containing \bquark or \cquark
quarks. The detector includes a high-precision tracking system
consisting of a silicon-strip vertex detector surrounding the $pp$
interaction region~\cite{LHCb-DP-2014-001},
a large-area silicon-strip detector located
upstream of a dipole magnet with a bending power of about
$4{\mathrm{\,Tm}}$, and three stations of silicon-strip detectors and straw
drift tubes~\cite{LHCb-DP-2013-003,LHCb-DP-2017-001}
placed downstream of the magnet.
The tracking system provides a measurement of the momentum, $p$, of charged particles with
a relative uncertainty that varies from 0.5\% at low momentum to 1.0\% at 200\gevc.
The minimum distance of a track to a primary $pp$-collision vertex (PV), the impact parameter (IP), 
is measured with a resolution of $(15+29/\pt)\mum$,
where \pt is the component of the momentum transverse to the beam, in\,\gevc.
Different types of charged hadrons are distinguished using information
from two ring-imaging Cherenkov detectors\cite{LHCb-DP-2012-003}. 
The online event selection is performed by a trigger~\cite{LHCb-DP-2012-004}, 
which consists of a hardware stage, based on information from the calorimeter and muon
systems, followed by a software stage, which applies a full event
reconstruction.

Simulated samples are required to develop the event selection and
to estimate the detector acceptance and the
efficiency of the imposed selection requirements.
Simulated $pp$ collisions are generated using
\pythia~\cite{Sjostrand:2007gs,*Sjostrand:2006za} 
with a specific \lhcb configuration~\cite{LHCb-PROC-2010-056}.  
A dedicated generator, \genxicctwo~\cite{Chang:2009va},
is used to simulate the doubly charmed baryon production.
Decays of unstable particles
are described by \evtgen~\cite{Lange:2001uf}, in which final-state
radiation is generated using \photos~\cite{davidson2015photos}. The
interaction of the generated particles with the detector, and its response,
are implemented using the \geant
toolkit~\cite{Allison:2006ve, *Agostinelli:2002hh} as described in
Ref.~\cite{LHCb-PROC-2011-006}. The \Omegacc\to\Xicp\Km\pip decay is assumed to proceed according to a uniform phase-space model. The \Omegacc baryon
and \Xiccpp baryon are assumed to have no polarization.
Unless otherwise stated,
simulated events are generated with an \Omegacc (\Xiccpp) mass of 3738\mevcc (3621\mevcc) 
and a lifetime of 160\fs (256\fs).

\section{Reconstruction and selection}%
\label{sec:reconstruction_and_selection}

The \Omegacc signal mode is reconstructed by combining a \Xicp candidate with kaon and pion candidates coming from the same vertex. The \Xicp candidates are firstly formed by combining three tracks originating from the same vertex, displaced with respect to the PV; at least one track is required to satisfy an inclusive software trigger based on a multivariate classifier~\cite{pmlr-v14-gulin11a, LHCb-PROC-2015-018}, and the three tracks must satisfy particle identification (PID) requirements to be compatible with a $\proton\Km\pip$ hypothesis. Then the \Xicp candidates with good vertex quality and invariant mass within the region of 2450 to 2486 \mevcc are combined with two extra tracks, identified as \Km and \pip , to reconstruct a \Omegacc candidate. The \Xicp mass region is defined as $2468\pm18 \mevcc$ where the mean value is the known \Xicp mass~\cite{PDG2020} and 
the width is corresponding to three
times the mass resolution.

To improve further the \Omegacc signal purity, a multivariate classifier based on a boosted decision tree (BDT)~\cite{Breiman,Hocker:2007ht,*TMVA4} is developed to suppress combinatorial background. 
The classifier is trained using simulated \Omegacc events as signal and wrong-sign $\Xicp\Km\pim$ combinations in data with mass 
in the interval 3600 to 4000\mevcc to represent background.

For \mbox{selection A}, no specific trigger requirement is applied. A multivariate selection is trained with two sets of variables which show good discrimination between \Omegacc signal and background. The first set contains variables related to the reconstructed \Omegacc candidates, including the \Omegacc decay vertex-fit quality, such as \chisqip, the pointing angle and the flight-distance \chisq. Here \chisqip is the difference in \chisq of the PV reconstructed with and without the \Omegacc candidate, the pointing angle is the three-dimensional angle between the \Omegacc
candidate momentum direction and the vector joining the PV and the reconstructed \Omegacc decay vertex, while the flight-distance \chisq is defined as the \chisq of the hypothesis that the decay vertex of the candidate coincides with its associated PV. The second set adds variables related to the decay products (\proton, \Km and \pip from the \Xicp decay, and \Km and \pip from the \Omegacc decay), including  momentum,  transverse momentum,  \chisqip and  PID variables.

The threshold of the multivariate output is determined by maximising 
the figure of merit $\varepsilon/\left(5/2 + \sqrt{N_B}\right)$~\cite{Punzi:2003bu}, where $\varepsilon$ is the estimated MVA selection efficiency, 
$5/2$ corresponds to 5 standard deviations in a Gaussian significance test, 
and $N_B$ is the expected number of background candidates
in the signal region, estimated with the wrong-sign $\Xicp\Km\pim$ combinations in the mass region of $\pm12.5\mevcc$ around the  \Omegacc mass of 3738\mevcc used in the simulation, taking into account the difference of the background level for the signal sample and the wrong-sign sample.

After the multivariate selection, the reconstructed \Omegacc candidates could suffer from background from candidates reconstructed with clone tracks, i.e. reconstructed tracks sharing a large portion of their detector hits. Clone tracks could be included in a \Omegacc candidate, when one is used for the \pip candidate from the \Xicp decay and its clone for the \pip candidate from the \Omegacc decay. To avoid that, candidates with the angle between each pair of identically charged tracks smaller than 0.5~mrad are removed. The \Omegacc candidates could also be formed by the same five final tracks but with two tracks interchanged, \eg the \Km (\pip) candidate from the \Xicp decay is swapped with the \Km (\pip) candidate from the \Omegacc decay. In this case, only one  candidate is chosen randomly.

For \mbox{selection B},
the multivariate selection is similar to \mbox{selection A} except that the 
PID variables of the \Km and \pip candidates from the \Omegacc decay are not used in the training to ease the efficiency determination. Furthermore, an additional hardware trigger requirement is imposed on candidates for both the signal and the normalisation modes to minimise differences between data and simulation. 
The data sets are split into two disjoint subsamples. One subsample is triggered on signals associated with one of the reconstructed \Xicp candidates with high transverse energy deposits in the calorimeters (TOS), and the other is triggered on signals exclusively unrelated to the \Omegacc candidate (exTIS).

The reconstruction and selection requirements of the \Xiccpp normalisation mode are similar to those in the \Xiccp search~\cite{LHCb-PAPER-2019-029,LHCb-PAPER-2019-035}.
Both \Omegacc and \Xiccpp candidates are required to be in the fiducial region of
rapidity $2.0<y<4.5$ and transverse momentum $4<\pt<15\gevc$.

\section{Yield measurements}
\label{sec:yield}

\begin{figure}[tb]
  \centering
  \includegraphics[width=0.70\linewidth]{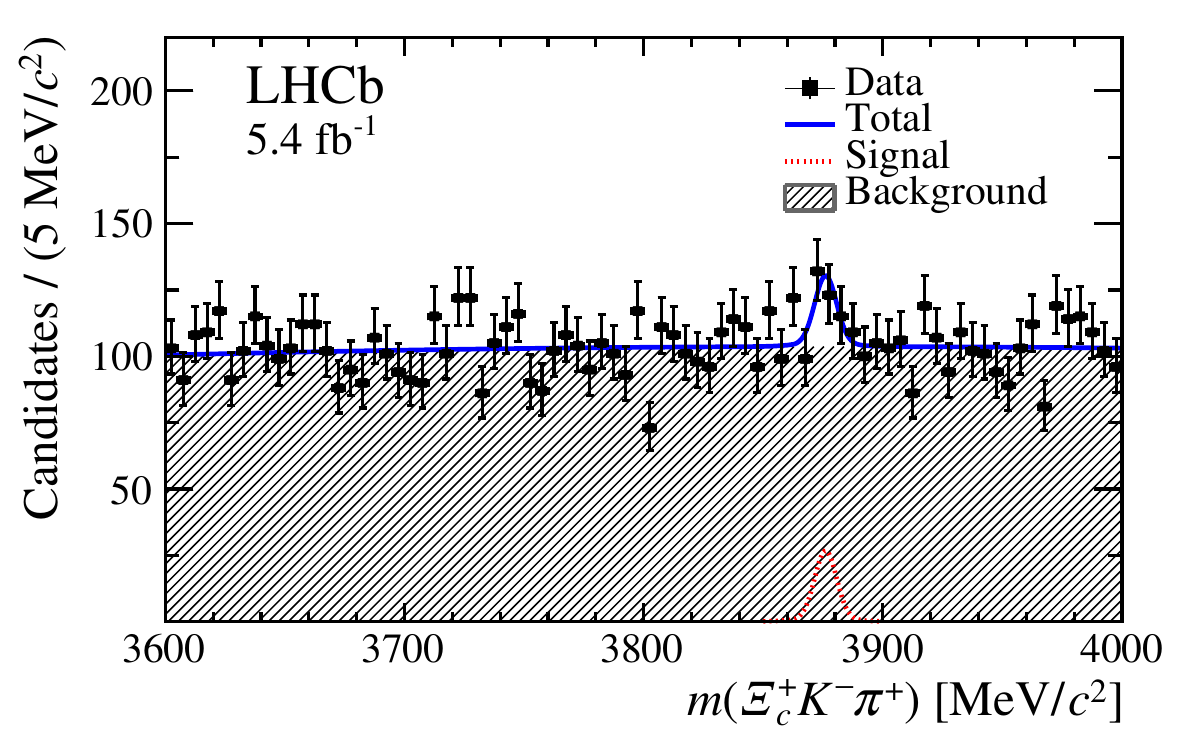}
  \caption{Invariant mass $m(\Xicp\Km\pip)$ distribution of selected \Omegacc candidates from (black points) selection A, with (blue solid line) the fit with the largest local significance at the mass of 3876\mevcc superimposed.}
  \label{fig:fit_shape}
\end{figure}

After applying \mbox{selection A} to the full data sample,  
the invariant mass distribution $m(\Xicp\Km\pip)$ of selected $\Omegacc$ candidates is shown in Fig.~\ref{fig:fit_shape}. 
To improve the mass resolution, 
the variable 
 $m(\Xicp\Km\pip)$ is defined as the difference of the reconstructed mass of the \Omegacc and \Xicp candidates plus the known \Xicp mass~\cite{PDG2020}.
The $m(\Xicp\Km\pip)$ distribution is fitted with a sum of signal and background components, where the signal component is described by the sum of two Crystal Ball functions~\cite{Skwarnicki:1986xj} and the background component by a second-order Chebyshev function. The parameters of the signal shape are fixed from simulation, where the width is found to be around 5.5\mevcc. The parameters of background shape are obtained from a fit to the wrong-sign $\Xicp\Km\pim$ combinations. An unbinned maximum likelihood fit is performed with the peak position varied in steps of \\
2 \mevcc, and the largest signal contribution is found for an \Omegacc mass of 3876\mevcc.

\begin{figure}[tb]
  \centering
  \includegraphics[width=0.7\linewidth]{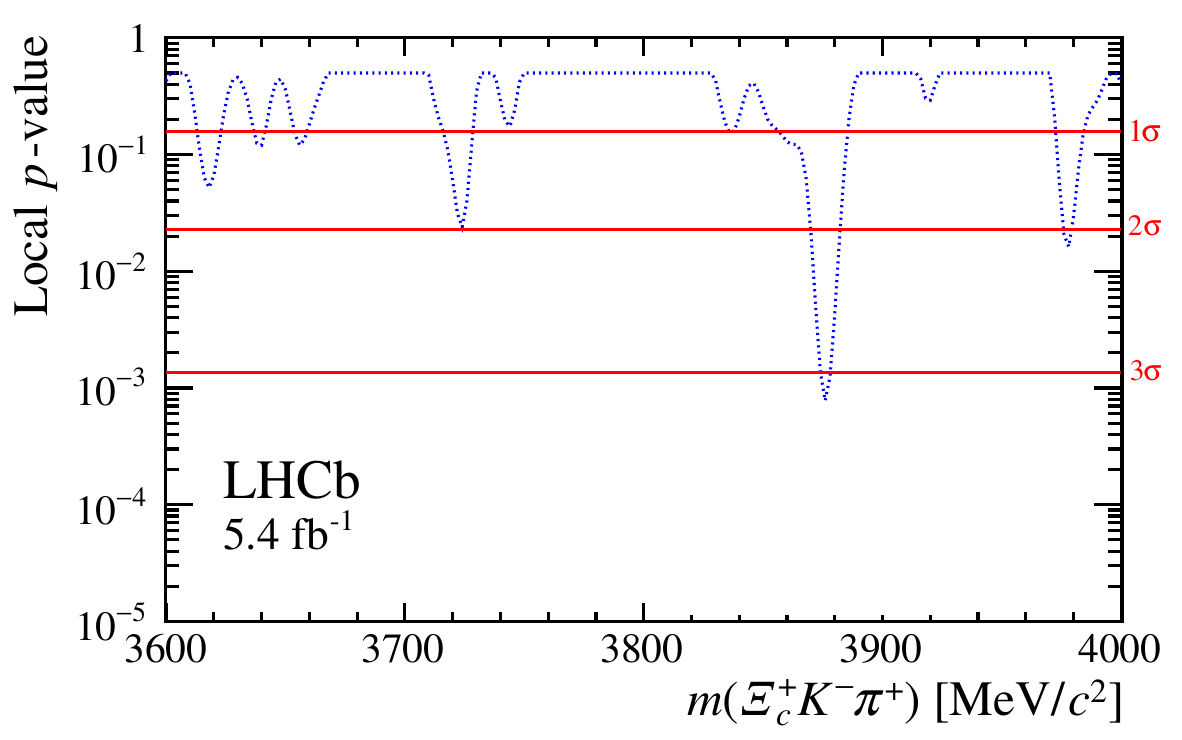}
  \caption{Local $p$-value at different $m(\Omegacc)$ values
  evaluated with the likelihood ratio test. Lines indicating one, two and three standard deviations ($\sigma$)  of local significance are also shown.}
  \label{fig:pvalue}
\end{figure}

\begin{figure}[tb]
  \centering
  \includegraphics[width=0.70\linewidth]{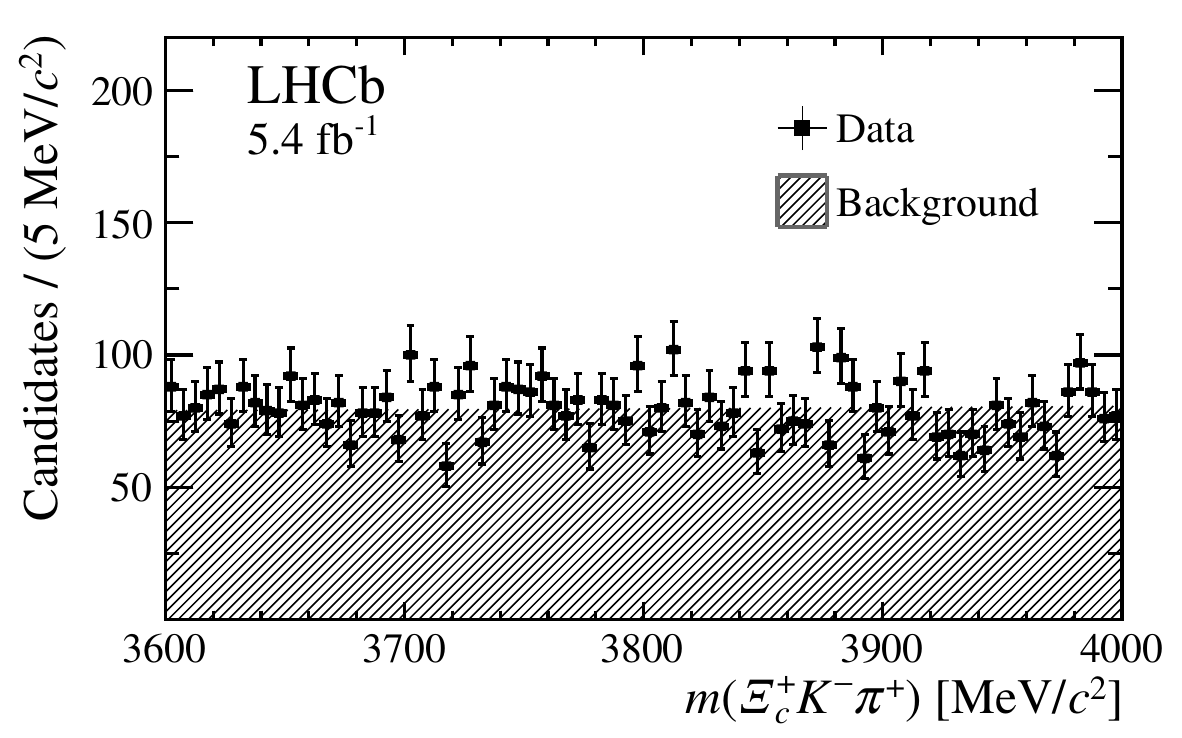}
  \caption{Invariant mass $m(\Xicp\Km\pip)$ distribution of selected \Omegacc candidates (black points) with selection B, only background fit is shown.}
  \label{fig:selectionB_fit}
\end{figure}

The local significance of the signal peak is quantified with a $p$-value, which is calculated as the likelihood ratio of the background plus signal hypothesis and the background-only hypothesis~\cite{Wilks:1938dza,Narsky:2000}.
The local $p$-value is plotted in Fig.~\ref{fig:pvalue} as a function of mass, $m(\Xicp\Km\pip)$, showing a dip around 3876\mevcc, which has the largest local significance, corresponding to $3.2$
standard deviations. The global significance is evaluated with pseudoexperiments, by taking into account the look-elsewhere effect~\cite{Gross:2010qma} in the mass range from 3600\mevcc to 4000\mevcc, and is estimated to be $1.8$ standard deviations.
As no excess above 3 standard deviations is observed,
upper limits on the production ratios are set by using \mbox{selection B}. The invariant mass distribution of \Omegacc candidates is shown in Fig.~\ref{fig:selectionB_fit} with the fit under the background-only hypothesis.

The measured production ratio is a function of single-event sensitivity $\alpha$ and $N_{\text{sig}}$, as shown in Eq.~\ref{eq:alphaN}. The parameter $\alpha$ is calculated using the yield of the normalisation mode $N_{\text{norm}}$ multiplied by the efficiency ratio between the normalisation and signal modes, while $N_{\text{sig}}$ is extracted by fitting the data of the signal mode.

\begin{figure}[tb]
  \centering
  \includegraphics[width=0.98\linewidth]{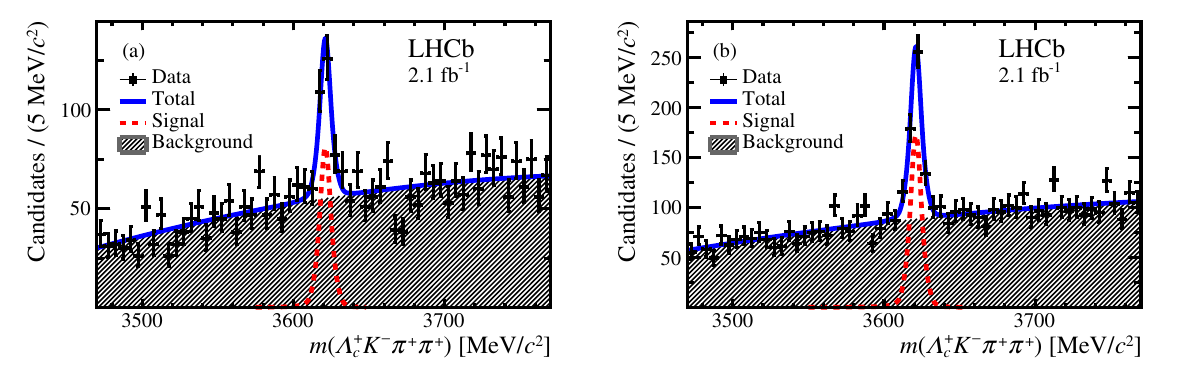}
  \caption{Distribution of invariant mass $m(\Lc\Km\pip\pip)$ for selected $\Xiccpp$ candidates in different categories:  (a)~triggered by one of the \Lc decay products and (b)~triggered exclusively by particles unrelated to the \Xiccpp decay products, in the 2018 data set. The fit results are superimposed.}
  \label{fig:yield_xicc}
\end{figure}

The $\Xiccpp$ yields, $N_{\text{norm}}$, are determined by performing an extended unbinned maximum likelihood fit to the invariant mass in the two trigger categories. The invariant mass distribution $m(\Lc\Km\pip\pip)$ is defined as the difference of the reconstructed mass of the \Xiccpp and \Lc candidates plus the known \Lc mass~\cite{PDG2020}.
For illustration,
the $m(\Lc\Km\pip\pip)$ distributions for the 2018 data set
are shown in Fig.~\ref{fig:yield_xicc} together with  the associated fit projections. 
The mass shapes of the normalisation mode are a sum of a Gaussian function and a modified Gaussian function with power-law tails on both sides for signal and a second-order Chebyshev polynomial for background, which is the same as used in the \Xiccp search~\cite{LHCb-PAPER-2019-035}.
The \Xiccpp yields are summarised in Table~\ref{tab:yield_summary}, where the TOS refers to the trigger on signal and the exTIS refers to exclusive trigger independently of signal. 

\begin{table}[bt]
  \centering
  \caption{Signal yields for the $\Xiccpp\to\Lc\Km\pip\pip$ normalisation mode $N_{\text{norm}}$ for both trigger categories and different data-taking periods with the corresponding integrated luminosity $\mathcal{L}$. The uncertainties are statistical only.}
  \label{tab:yield_summary}
  \begin{tabular}{cccc}
    \toprule
    \multirow{2}*{Year} & \multirow{2}*{$\mathcal{L}$ $[\invfb]$} &\multicolumn{2}{c}{$N_{\text{norm}}$} \\ 
 \cline{3-4}
 ~ & ~ & \multirow{1.2}*{TOS} & \multirow{1.2}*{exTIS}\\
    \midrule
    2016&   1.7&  $126\pm21$   & $165\pm23$ \\
    2017&   1.6&  $145\pm21$   & $255\pm26$ \\
    2018&   2.1&  $164\pm21$  & $349\pm30$ \\
    \bottomrule
  \end{tabular}
\end{table}

\section{Efficiency ratio estimation}%
\label{sec:efficiency_ratio}
The efficiency ratio between the \Xiccpp mode and \Omegacc mode, defined as $\eps_{\rm norm}/\eps_{\rm sig}$, is determined from simulation.
The distributions of the transverse momentum, rapidity of the doubly charmed baryons, and the event multiplicity in simulated samples are weighted according to the differences between simulation and data seen for the \Xiccpp baryon.  The Dalitz distributions of the simulated intermediate \Xicp (\Lc)  states are corrected to match those in data. 
The tracking and PID efficiencies for both normalisation and signal modes are corrected using calibration data samples~\cite{LHCb-DP-2013-002,LHCb-PUB-2016-021,LHCb-DP-2018-001}. 
The efficiency ratios of both trigger categories for different data-taking periods are summarised in Table~\ref{tab:eff_ratio}. Since there is an additional track in the \Xiccpp decay when compared to the \Omegacc decay, the reconstruction and selection efficiency of \Xiccpp candidates is significantly lower. The increase in the efficiency ratio for the 2017 and 2018 data is due to the optimisation of the \Xiccpp online selection, following the observation of the \Xiccpp baryon~\cite{LHCb-PAPER-2017-018}.

\begin{table}[tb]
  \centering
  \caption{Efficiency ratios $\eps_{\rm norm}/\eps_{\rm sig}$ between  normalisation and signal modes for both trigger categories for 
  different data-taking periods, where the TOS refers to the trigger on signal and the exTIS refers to exclusive trigger independently of signal. The uncertainties are statistical only.}
\label{tab:eff_ratio}
\begin{tabular}{ccc} 
\toprule 
\multirow{2}*{Year} & \multicolumn{2}{c}{$\eps_{\rm norm}/\eps_{\rm sig}$} \\ 
 \cline{2-3}
 ~ & \multirow{1.2}*{TOS} & \multirow{1.2}*{exTIS}\\
\midrule 
2016& 0.32 $\pm$ 0.03 & 0.28 $\pm$ 0.02\\ 
2017& 0.55 $\pm$ 0.03 & 0.71 $\pm$ 0.02\\ 
2018& 0.61 $\pm$ 0.04 & 0.69 $\pm$ 0.02\\ 
\bottomrule 
\end{tabular} 
\end{table}

In order to take into account the dependence of the selection efficiency upon the unknown value of the \Omegacc lifetime, simulated \Omegacc events
are weighted to reproduce different exponential decay time distributions corresponding to lifetimes of 40, 80, 120, 160, and 200\fs. This method is used to estimate the change in the efficiency. 
The single-event sensitivities are calculated by the ratio of \Xiccpp efficiency to the \Omegacc efficiency with different lifetime hypotheses, as shown in Tables~\ref{tab:alpha_TOS} and~\ref{tab:alpha_TIS}, for both trigger categories.

\begin{table}[tb]
  \centering
  \caption{Single-event sensitivity $\alpha(\Xiccpp)$ [$10^{-2}$] of the \Xiccpp normalisation mode triggered by one of the \Xicp (\Lc) products for different lifetime hypotheses of the $\Omegacc$ baryon for different data-taking periods.
  The uncertainties are due to the limited size of the simulated samples
  and the statistical uncertainties on the measured \Xiccpp baryon yields.}
\label{tab:alpha_TOS}
\begin{tabular}{c rrrrr} 
\toprule 
\multirow{2}*{Year} & \multicolumn{5}{c}{$\alpha$ [$10^{-2}$]} \\ 
 \cline{2-6}
~ & \multirow{1.2}*{$\tau=40\fs$} & \multirow{1.2}*{$\tau=80\fs$} & \multirow{1.2}*{$\tau=120\fs$} & 
\multirow{1.2}*{$\tau=160\fs$} & \multirow{1.2}*{$\tau=200\fs$}\\ 
\midrule 
2016& 0.86 $\pm$ 0.17 & 0.46 $\pm$ 0.09 & 0.32 $\pm$ 0.06 & 0.25 $\pm$ 0.05 & 0.22 $\pm$ 0.04\\ 
2017& 1.29 $\pm$ 0.20 & 0.69 $\pm$ 0.11 & 0.48 $\pm$ 0.07 & 0.38 $\pm$ 0.06 & 0.33 $\pm$ 0.05\\ 
2018& 1.26 $\pm$ 0.18 & 0.67 $\pm$ 0.10 & 0.47 $\pm$ 0.07 & 0.37 $\pm$ 0.05 & 0.32 $\pm$ 0.05\\ 
\bottomrule 
\end{tabular} \end{table}

\begin{table}[!tb]
  \centering
  \caption{Single-event sensitivity $\alpha(\Xiccpp)$ [$10^{-2}$] of the \Xiccpp normalisation mode triggered exclusively by particles unrelated to the \Omegacc (\Xiccpp) decay products for different lifetime hypotheses of the $\Omegacc$ baryon in the different data-taking periods.
  The uncertainties are due to the limited size of the simulated samples
  and the statistical uncertainty on the measured \Xiccpp baryon yield.}
\label{tab:alpha_TIS}
\begin{tabular}{c rrrrr} 
\toprule 
\multirow{2}*{Year} & \multicolumn{5}{c}{$\alpha$ [$10^{-2}$]} \\ 
 \cline{2-6}
~ & \multirow{1.2}*{$\tau=40\fs$} & \multirow{1.2}*{$\tau=80\fs$} & \multirow{1.2}*{$\tau=120\fs$} & 
\multirow{1.2}*{$\tau=160\fs$} & \multirow{1.2}*{$\tau=200\fs$}\\
\midrule 
2016& 0.71 $\pm$ 0.11 & 0.35 $\pm$ 0.06 & 0.22 $\pm$ 0.04 & 0.17 $\pm$ 0.03 & 0.14 $\pm$ 0.02\\ 
2017& 1.16 $\pm$ 0.12 & 0.57 $\pm$ 0.06 & 0.37 $\pm$ 0.04 & 0.28 $\pm$ 0.03 & 0.23 $\pm$ 0.02\\ 
2018& 0.82 $\pm$ 0.08 & 0.41 $\pm$ 0.04 & 0.26 $\pm$ 0.02 & 0.20 $\pm$ 0.02 & 0.17 $\pm$ 0.02\\ 
\bottomrule 
\end{tabular} \end{table}

The \Omegacc mass is also unknown. To test the effects of different mass hypotheses,
two simulated samples are generated with $m(\Omegacc)=3638\mevcc$ and $m(\Omegacc)=3838\mevcc$.
These samples are used to weight the \pt distributions of final states in the \Omegacc decay
to match those in the other mass hypotheses, and the efficiency is recalculated with the weighted samples. When varying the \Omegacc mass, it is found that the efficiency is constant; therefore, the \Omegacc mass dependence is neglected in the evaluation of the single-event sensitivities.

\section{Systematic uncertainties}
The sources of systematic uncertainties on the production ratio $R$ are listed in Table~\ref{tab:sys_fit}, where individual sources are assumed to be independent and summed in quadrature to compute the total systematic uncertainty.  

The choice of the mass models used to fit the invariant mass distribution affects the normalisation yields and therefore affects the calculation of single-event sensitivities. The related systematic uncertainty is studied by using alternative functions to describe the signal and background shapes of the \Xiccpp mode.
The sum of two Gaussian functions is chosen as an alternative signal model and a second-order polynomial function is chosen to substitute the background model. The difference in the signal yields obtained by changing models is assigned as the systematic uncertainty.

The systematic uncertainty associated with the trigger efficiency is evaluated using a tag-and-probe method~\cite{LHCb-DP-2012-004}.
The size of the normalisation sample is insufficient to derive this systematic uncertainty. Instead, $b$-flavoured hadrons decaying with similar final-state topologies are used. 
For the TOS category, 
 \Lb\to\Lc\pip\pim\pim and
\Lb\to\Lc\pim candidates can be 
triggered by the energy deposit in the calorimeter by one of the \Lc decay products, which are similar to 
the $\Xiccpp \to \Lc\Km\pip\pip$ and $\Omegacc \to \Xicp\Km\pip$ decays. 
The efficiency ratio of these two \Lb modes is estimated and the difference of the ratio between data and simulation is assigned as a systematic uncertainty.  
For the exTIS category, the \Bc\to\jpsi\pip decay, which has two heavy-flavour particles ($b$- and $c$-hadrons) and is similar to the signal topology, is used to study the trigger efficiency with particle candidates that are independent and unrelated to the signal. The systematic uncertainty for the exTIS trigger category is assigned as the difference in the efficiency ratio of \Lb\to\Lc\pip\pim\pim mode to \Bc\to\jpsi\pip mode in data and in simulation.

The tracking efficiency is corrected with calibration data samples~\cite{LHCb-DP-2013-002}, and is affected by three sources of systematic uncertainties. 
First, the inaccuracy of the simulation in terms of detector occupancy, which is assigned as 1.5\% and 2.5\% for kaons and pions,  does not cancel in the ratio. An additional systematic uncertainty arises from the calibration method which provides a 0.8\% uncertainty per track~\cite{LHCb-DP-2013-002}. The third uncertainty is due to the limited size of the calibration samples and studied by pseudoexperiments. The tracking efficiency is corrected by the pseudoexperiments and the Gaussian width of the newly obtained distribution of the efficiency ratio is assigned as the systematic uncertainty.

The PID efficiency is determined in intervals of particle momentum, 
pseudorapidity and event multiplicity using calibration data samples. The corresponding sources of systematic uncertainty are due to the limited size of the calibration samples and the binning scheme used. To study their effects, a large number of pseudoexperiments are performed, and the binning scheme is varied.

The \Xiccpp lifetime is measured with limited precision, $\ensuremath{256\,^{+24}_{-22}\,{\rm(stat)\,}\pm
14\,{\rm(syst)}\,\fs}$~\cite{LHCb-PAPER-2018-019}, which is propagated to the systematic uncertainty in the efficiency.

As the agreeemnt between data and simulation is limited, a difference of 5.0\% is found among different periods of data-taking, which is taken as systematic uncertainty.

 \begin{table}[tp]
    \centering
      \caption{Systematic uncertainties on the production ratio $R$.}
  \label{tab:sys_fit}
  \begin{tabular}{lr}
    \toprule
    Source & $R$~[\%] \\
    \midrule
    Fit model             & 3.5 \\
    Hardware trigger & 11.2  \\
    Tracking             & 2.7  \\
    PID   & 0.9  \\
    \Xiccpp lifetime &  12.0\\
    Simulation/data difference                  & 5.0 \\
       \midrule
    Total                        & 17.7 \\
    \bottomrule
  \end{tabular}
\end{table}

\section{Results}%
\label{sec:results}
Upper limits on the production ratio $R$ are set with a simultaneous fit to the $m(\Xicp\Km\pip)$ distributions of different trigger categories for all the data sets from 2016 to 2018,
following the strategy described in Sec.~\ref{sec:yield} for the normalisation mode.
The upper limit values are calculated by setting different \Omegacc mass hypotheses in the fit within the $m(\Xicp\Km\pip)$ mass range from 3600 to 4000\mevcc with a step of 2\mevcc, for five different lifetime hypotheses, 40, 80, 120, 160, and 200\fs.

\begin{figure}[tb]
  \centering
  \includegraphics[width=0.70\linewidth]{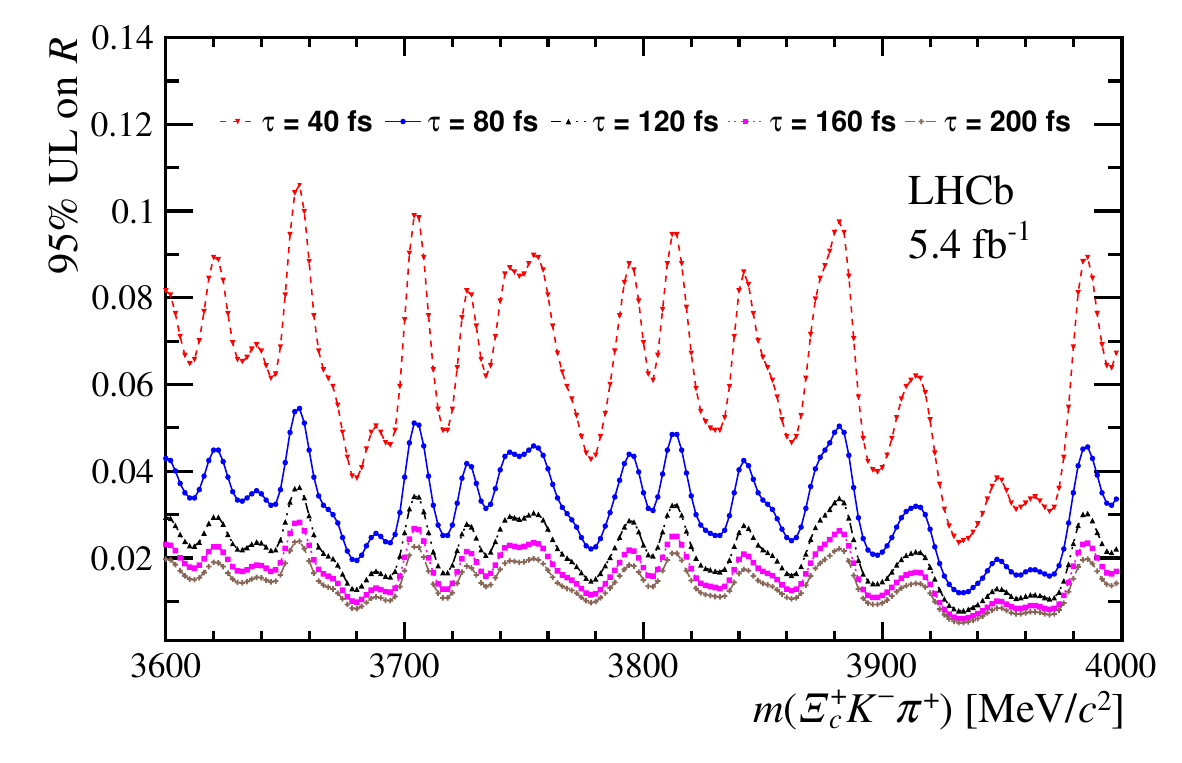}
  \caption{Upper limits on the production ratio $R$
  at 95\% credibility level as a function of $m(\Xicp\Km\pip)$ at
  $\sqrt{s}=13\tev$, 
  for five $\Omegacc$ lifetime hypotheses.}
  \label{fig:mass_scan_tau_run2}
\end{figure}

For each \Omegacc mass and lifetime hypothesis,  
the likelihood profile 
is determined as a function of $R$.
It is then convolved with a Gaussian distribution whose width is equal to the square root of the quadratic combination of the 
statistical and systematic uncertainties on the single-event sensitivity.
The upper limit at 95\% credibility level is defined as
the value of $R$ at which the integral of the profile likelihood equals 95\% of the total area. Figure~\ref{fig:mass_scan_tau_run2} 
shows the 95\% credibility level upper limits at different mass hypotheses for five different lifetimes. The upper limits on $R$ decrease when increasing the \Omegacc lifetime. Considering the whole explored mass range, the highest upper limit on $R$ is $0.11$  obtained under lifetime hypothesis of 40 \fs while the lowest is $0.5 \times 10^{-2}$ obtained under lifetime hypothesis of 200 \fs. 

\section{Conclusion}%
\label{sec:conclusion}

A search for the \Omegacc baryon through the \Xicp\Km\pip decay is performed, using $pp$ collision data collected by the \lhcb experiment
from 2016 to 2018 at a centre-of-mass energy of 13\tev,  corresponding to an integrated luminosity of 5.4\invfb.
No significant signal is observed in the mass range of 3.6 to 4.0\gevcc.
Upper limits are set at $95\%$ credibility level
on the ratio of the \Omegacc production cross-section times
the branching fraction to that of the \Xiccpp baryon as a function of the \Omegacc mass and for different lifetime hypotheses, 
in the rapidity range of 2.0 to 4.5 and the transverse momentum range of 4 to 15\gevc.
The upper limits are set assuming that the \Omegacc\to\Xicp\Km\pip decay proceeds according to a uniform phase-space model. 
The upper limits depend strongly on the mass and lifetime hypotheses of the \Omegacc, and vary from $1.1 \times 10^{-1}$ to $0.5 \times 10^{-2}$ for 40\fs to 200\fs, respectively.
Future searches by the \lhcb experiment with upgraded detectors, improved trigger conditions, additional
$\Omegacc$ decay modes, and larger data samples will further increase the \Omegacc signal sensitivity.

\section*{Acknowledgements}
\noindent 
We express our gratitude to our colleagues in the CERN
accelerator departments for the excellent performance of the LHC. We
thank the technical and administrative staff at the LHCb
institutes.
We acknowledge support from CERN and from the national agencies:
CAPES, CNPq, FAPERJ and FINEP (Brazil); 
MOST and NSFC (China); 
CNRS/IN2P3 (France); 
BMBF, DFG and MPG (Germany); 
INFN (Italy); 
NWO (Netherlands); 
MNiSW and NCN (Poland); 
MEN/IFA (Romania); 
MSHE (Russia); 
MICINN (Spain); 
SNSF and SER (Switzerland); 
NASU (Ukraine); 
STFC (United Kingdom); 
DOE NP and NSF (USA).
We acknowledge the computing resources that are provided by CERN, IN2P3
(France), KIT and DESY (Germany), INFN (Italy), SURF (Netherlands),
PIC (Spain), GridPP (United Kingdom), RRCKI and Yandex
LLC (Russia), CSCS (Switzerland), IFIN-HH (Romania), CBPF (Brazil),
PL-GRID (Poland) and NERSC (USA).
We are indebted to the communities behind the multiple open-source
software packages on which we depend.
Individual groups or members have received support from
ARC and ARDC (Australia);
AvH Foundation (Germany);
EPLANET, Marie Sk\l{}odowska-Curie Actions and ERC (European Union);
A*MIDEX, ANR, IPhU and Labex P2IO, and R\'{e}gion Auvergne-Rh\^{o}ne-Alpes (France);
Key Research Program of Frontier Sciences of CAS, CAS PIFI, CAS CCEPP, 
Fundamental Research Funds for the Central Universities, 
and Sci. \& Tech. Program of Guangzhou (China);
RFBR, RSF and Yandex LLC (Russia);
GVA, XuntaGal and GENCAT (Spain);
the Leverhulme Trust, the Royal Society
 and UKRI (United Kingdom).

\clearpage
\addcontentsline{toc}{section}{References}
\bibliographystyle{LHCb}
\bibliography{main,standard,LHCb-PAPER,LHCb-CONF,LHCb-DP,LHCb-TDR,new_Bib}
\clearpage
 
\centerline
{\large\bf LHCb collaboration}
\begin
{flushleft}
\small
R.~Aaij$^{32}$,
C.~Abell{\'a}n~Beteta$^{50}$,
T.~Ackernley$^{60}$,
B.~Adeva$^{46}$,
M.~Adinolfi$^{54}$,
H.~Afsharnia$^{9}$,
C.A.~Aidala$^{86}$,
S.~Aiola$^{25}$,
Z.~Ajaltouni$^{9}$,
S.~Akar$^{65}$
% ...

{\footnotesize \it

% ...
$^{9}$Universit{\'e} Clermont Auvergne, CNRS/IN2P3, LPC, Clermont-Ferrand, France\\
% ...
$^{25}$INFN Sezione di Milano, Milano, Italy\\
% ...
$^{32}$Nikhef National Institute for Subatomic Physics, Amsterdam, Netherlands\\
% ...
$^{46}$Instituto Galego de F{\'\i}sica de Altas Enerx{\'\i}as (IGFAE), Universidade de Santiago de Compostela, Santiago de Compostela, Spain\\
% ...
$^{50}$Physik-Institut, Universit{\"a}t Z{\"u}rich, Z{\"u}rich, Switzerland\\
% ...
$^{54}$H.H. Wills Physics Laboratory, University of Bristol, Bristol, United Kingdom\\
% ...
$^{60}$Oliver Lodge Laboratory, University of Liverpool, Liverpool, United Kingdom\\
% ...
$^{65}$University of Cincinnati, Cincinnati, OH, United States\\
% ...
$^{68}$Syracuse University, Syracuse, NY, United States\\
% ...
$^{86}$University of Michigan, Ann Arbor, United States, associated to $^{68}$\\
\bigskip
% ...
\medskip
}
\end{flushleft}

%
%
%
%
%

\end{document}